\documentclass{elsart}
\usepackage{epsfig}
\usepackage{graphicx}
\usepackage[figuresright]{rotating}

\def\be{\begin{equation}}
\def\ee{\end{equation}}
\def\ba{\begin{array}}
\def\ea{\end{array}}

\begin{document}
\begin{frontmatter}

\title{\boldmath A study of charged $\kappa$ in
         $J/\psi \to K^{\pm} K_S \pi^{\mp} \pi^0$ }


\begin{small}

\begin{center}
\vspace{0.2cm}
BES Collaboration\\
\vspace{0.2cm}
M.~Ablikim$^{1}$,              J.~Z.~Bai$^{1}$,   Y.~Bai$^{1}$,
Y.~Ban$^{11}$,
X.~Cai$^{1}$,                  H.~F.~Chen$^{16}$,
H.~S.~Chen$^{1}$,              H.~X.~Chen$^{1}$, J.~C.~Chen$^{1}$,
Jin~Chen$^{1}$
% ...
\\
{\small\it
$^{1}$ Institute of High Energy Physics, Beijing 100049, People's Republic of China\\
% ...
$^{11}$ Peking University, Beijing 100871, People's Republic of China\\
% ...
$^{16}$ University of Science and Technology of China, Hefei 230026,
People's Republic of China\\
% ...
}
\end{center}
\end{small}

~~\\
\begin{flushleft}
\end{flushleft}

\begin{abstract}

  Based on $58 \times 10^6 $ $J/\psi$ events collected by BESII, the
  decay $J/\psi \to K^{\pm} K_S \pi^{\mp} \pi^0$ is studied. In the
  invariant mass spectrum recoiling against the charged
  $K^*(892)^{\pm}$, the charged $\kappa$ particle is found as a low
  mass enhancement.  If a Breit-Wigner function of constant
  width is used to parameterize the $\kappa$, its pole locates at $(849 \pm 77
  ^{+18}_{-14}) -i (256 \pm 40 ^{+46}_{-22})$ MeV/$c^2$.  Also in this
  channel, the decay $J/\psi \to K^*(892)^+ K^*(892)^-$ is observed
  for the first time. Its branching ratio is $(1.00 \pm 0.19
  ^{+0.11}_{-0.32}) \times 10^{-3}$.

\end{abstract}

\maketitle
\end{frontmatter}

\section{Introduction}
The $\sigma$ and $\kappa$ are controversial particles in hadron
spectroscopy. They were first found in the analysis of $\pi\pi$ and
$\pi K$ scattering data. Because the total phase shifts in the lower
mass region are much less than 180$^{\circ}$ and they do not fit into
ordinary $q \bar q$ meson nonets, they have been the subject of
violent debates. Refs.~(\cite{bes1s} - \cite{s4}) are some recent
analyses that support their existence.

Evidence for $\kappa$ particles comes from the study of production
processes and the re-analysis of $K \pi$ scattering data.  Evidence
for the $\kappa$ has been found in $D^+ \to K^- \pi^+
\pi^-$~\cite{s0k}, $J/\psi \to \bar{K}^*(892)^0 K^+
\pi^-$~\cite{d1,bes2k}, and $D^+ \to K^- \pi^+ \mu^+ \nu$~\cite{d2}.
A $K \pi$ s-wave component is found in $D^0 \to K^- K^+
\pi^0$~\cite{d3}, $D^+ \to K^- \pi^+ e^+ \nu_e$~\cite{d3b}, and $\tau
\to K_S \pi^- \nu_{\tau}$~\cite{d4}.  But no evidence for the
$\kappa$ in $D^0 \to K^- \pi^+ \pi^0$~\cite{d5} or for the charged
$\kappa$ in $D^0 \to K^- K^+ \pi^0$~\cite{d6} is seen.  The $\kappa$
is found in the phenomenological analysis of $K \pi$ scattering phase
shift data~(\cite{s4},~\cite{c1} - \cite{hanqing}).  However, some
theorists are not convinced by the evidence~(\cite{c8} - \cite{c11}).
The present status of the $\kappa$ is summarized by the Particle Data
Group PDG~\cite{pdg}.

The $\sigma$ and $\kappa$ particles have been studied with BES data,
where evidence for $\sigma$ and $\kappa$ particles is quite
clear~(\cite{bes1s} - \cite{bes2k}).  A neutral $\kappa$ was found in
the decay $J/\psi \to \bar K^*(892)^0 \kappa^0 \to \bar K^*(892)^0 K^+
\pi^-$~\cite{d1,bes2k}.  Because of isospin symmetry, if a neutral
$\kappa$ exists, a charged $\kappa$ should exist and could be produced
in $J/\psi \to \bar K^*(892)^{\pm} \kappa^{\mp}$.

In this study, we search for and study the charged $\kappa$ in $J/\psi
\to K^{\pm} K_S \pi^{\mp} \pi^0$.  Our analysis is based on 58 million
$J/\psi$ decays collected by BESII at the BEPC (Beijing Electron
Positron Collider).  BESII is a large solid-angle magnetic
spectrometer which is described in detail in Ref.~\cite{besd}.  The
momentum of charged particles is determined by a 40-layer cylindrical
main drift chamber (MDC). Particle identification is accomplished
using specific ionization ($dE/dx$) measurements in the MDC and
time-of-flight (TOF) information in a barrel-like array of 48
scintillation counters.  Outside of the TOF is a barrel shower counter
(BSC) which measures the energy and direction of photons.

\section{Event selection}
In the event selection, candidate tracks are required to have a good
track fit with the point of closest approach of the track to the beam
axis being within the interaction region of 2 cm in $r_{xy}$ and $\pm
20$ cm in $z$ (the beam direction), polar angles $\theta$ satisfying
$|\cos \theta|<0.80$, and transverse momenta $P_t> 50$ MeV/$c$.
Photons are required to be isolated from charged tracks, to come from
the interaction region, and have deposited energy in the BSC greater
than 40 MeV.  Events are required to have four good charged
tracks with total charge zero and at least two good photons.

For the $K_S$ reconstruction, we loop over all oppositely
charged pairs of tracks, assuming them to be pions, and fit them to
$K_S \to \pi^+ \pi^-$, which determines vertices and $\pi^+\pi^-$
invariant masses, $M_{\pi\pi}$, for the four possible combinations.
The combination with $M_{\pi\pi}$ closest to $M_{K_S}$ is selected and is
required to satisfy $|M_{\pi^+ \pi^-} - M_{K_S}| < 20$ MeV/$c^2$ and
have its decay vertex in the $xy$-plane satisfy $r_{xy}>0.008$ m.
After $K_S$ selection, the particle type of the remaining two tracks,
that is whether they are $K^+ \pi^-$ or $K^- \pi^+$, is decided by
selecting the combination with smallest $\chi^2_{TOF} +
\chi^2_{DEDX}$.

A four constraint (4C) kinematic fit is applied under the $K^{\pm}
\pi^{\mp} \pi^+ \pi^- \gamma \gamma$ hypothesis, and $\chi^2_{4C} <
20$ is required. Events with a $\gamma \gamma$ invariant mass
satisfying $|M_{\gamma \gamma} - M_{\pi^0}| < 40$ MeV/$c^2$ are fitted
with a 5C kinematic fit to $K^{\pm} \pi^{\mp} \pi^+ \pi^- \pi^0$ with
the two photons constrained to the $\pi^0$ mass, and events with
$\chi^2_{5C} < 50$ are selected.  The $\pi^{\mp} \pi^0$ mass
distribution is shown in Fig.~1(a), where the $\rho$ is clearly seen.
Decays with an intermediate $\rho$ are background,
and the requirement $|M_{\pi^{\mp} \pi^0} - M_{\rho}| >
100$ MeV/$c^2$ is applied to remove them. The requirements $| M_{K_S
  \pi^0} - 0.897 | > 40$ MeV/$c^2$ and $| M_{K^{\pm} \pi^{\mp}} -
0.897 | > 40$ MeV/$c^2$ are used to remove backgrounds from $J/\psi
\to K^*(892)^0 K^{\pm} \pi^{\mp}$ and $J/\psi \to K^*(892)^0 K_S
\pi^0$.  After these requirements, the combined $K^{\pm} \pi^0$ and
$K_S \pi^{\mp}$ mass distribution is shown in Fig.~1(b); the highest
narrow peak is charged $K^*(892)$. The $M_{K^{\pm}\pi^0}$ versus
$M_{K_S \pi^{\mp}}$ scatter plot is shown in Fig.~1(c).  There are
two clear bands, a vertical and horizontal band, which
correspond to $J/\psi \to K^*(892)^{\pm} K_S \pi^{\mp} $ and $J/\psi
\to K^*(892)^{\pm} K^{\mp} \pi^0$, respectively.  The requirements
$|M_{K^{\pm} \pi^0} - 0.892| < 80$ MeV/$c^2$ and $|M_{K_S \pi^{\pm}} -
0.892| < 80$ MeV/$c^2$ are imposed to select $J/\psi \to K^*(892)^{\pm} K_S
\pi^{\mp} $ and $J/\psi \to K^*(892)^{\pm} K^{\mp} \pi^0$ events,
respectively.

After the above selection, the combined $K_S \pi^{\mp}$ and $K^{\mp}
\pi^0$ invariant mass distribution recoiling against the
$K^*(892)^{\pm}$ is shown in Fig.~1(d). It is the sum of the
four decays listed in Table~1 with about 1000 events for each, as listed
in the table, and a total of 4121 events. 
In Fig.~1(d), a clear narrow peak at 892 MeV/$c^2$ and a wider peak at
about 1430 MeV/$c^2$ are seen. In addition, there is a broad low mass
enhancement just above threshold.  The spectrum is quite similar to
the spectrum of $K^+ \pi^-$ in the decay $J/\psi \to \bar{K}^*(892)^0
K^+ \pi^-$~\cite{bes2k}.  The biggest difference between them is that
the charged $K^*(892)$ peak is much larger than the neutral $K^*(892)$
peak.  The $K^*(892) \pi$ invariant mass distribution is shown in
Fig.~1(e), where there are indications of peaks at 1270 MeV/$c^2$ and
1400 MeV/$c^2$.  The resulting Dalitz plot is shown in Fig.~1(f). The
two diagonal bands correspond to the low mass enhancement combined
with the 892 MeV/$c^2$ peak and the peak around 1430 MeV/$c^2$ in the
$K \pi$ spectrum.

\begin{table}[htp]
\begin{center}
\doublerulesep 0pt
\renewcommand\arraystretch{1.1}
\begin{tabular}{|c|c|}
\hline

Channel & Events   \\
\hline
$J/\psi \to K^*(892)^+ K_S \pi^-  \to K^+ \pi^0 K_S \pi^-$  &  1023  \\
\hline
$J/\psi \to K^*(892)^- K_S \pi^+  \to K^- \pi^0 K_S \pi^+$  &  946  \\
\hline
$J/\psi \to K^*(892)^+ K^- \pi^0  \to K_S \pi^+ K^- \pi^0$  &  1055  \\
\hline
$J/\psi \to K^*(892)^- K^+ \pi^0  \to K_S \pi^- K^+ \pi^0$  &  1097  \\
\hline
\end {tabular}
\vspace{0.1in}
\caption {Four signal channels and number of events from each.}
\end{center}
\end{table}

\begin{figure}[htbp]
\begin{flushleft}
{\mbox{\epsfig{file=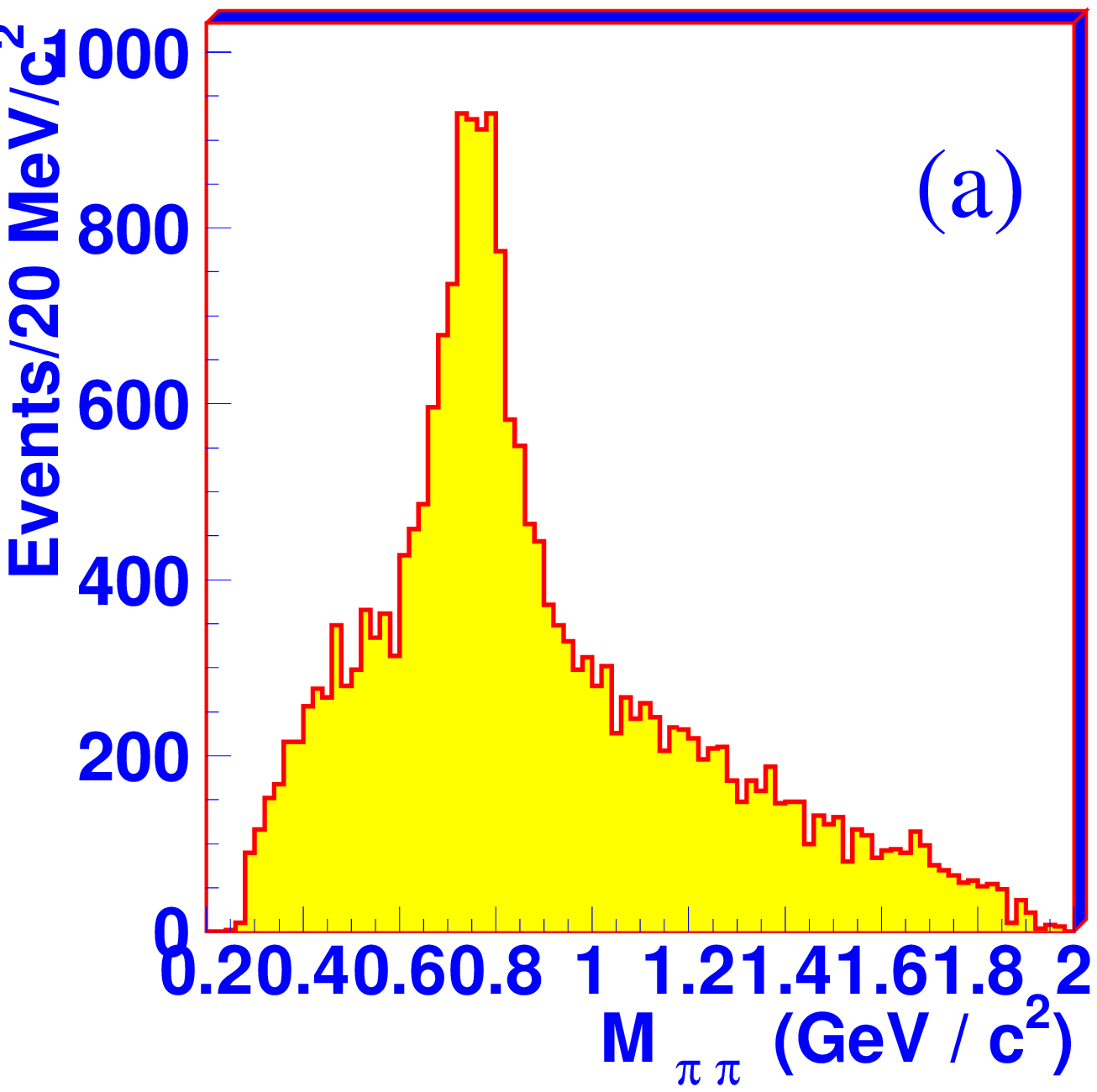,height=2.6in,width=2.6in}}}
\end{flushleft}
\vspace{-2.6in} 
\begin{flushright}
{\mbox{\epsfig{file=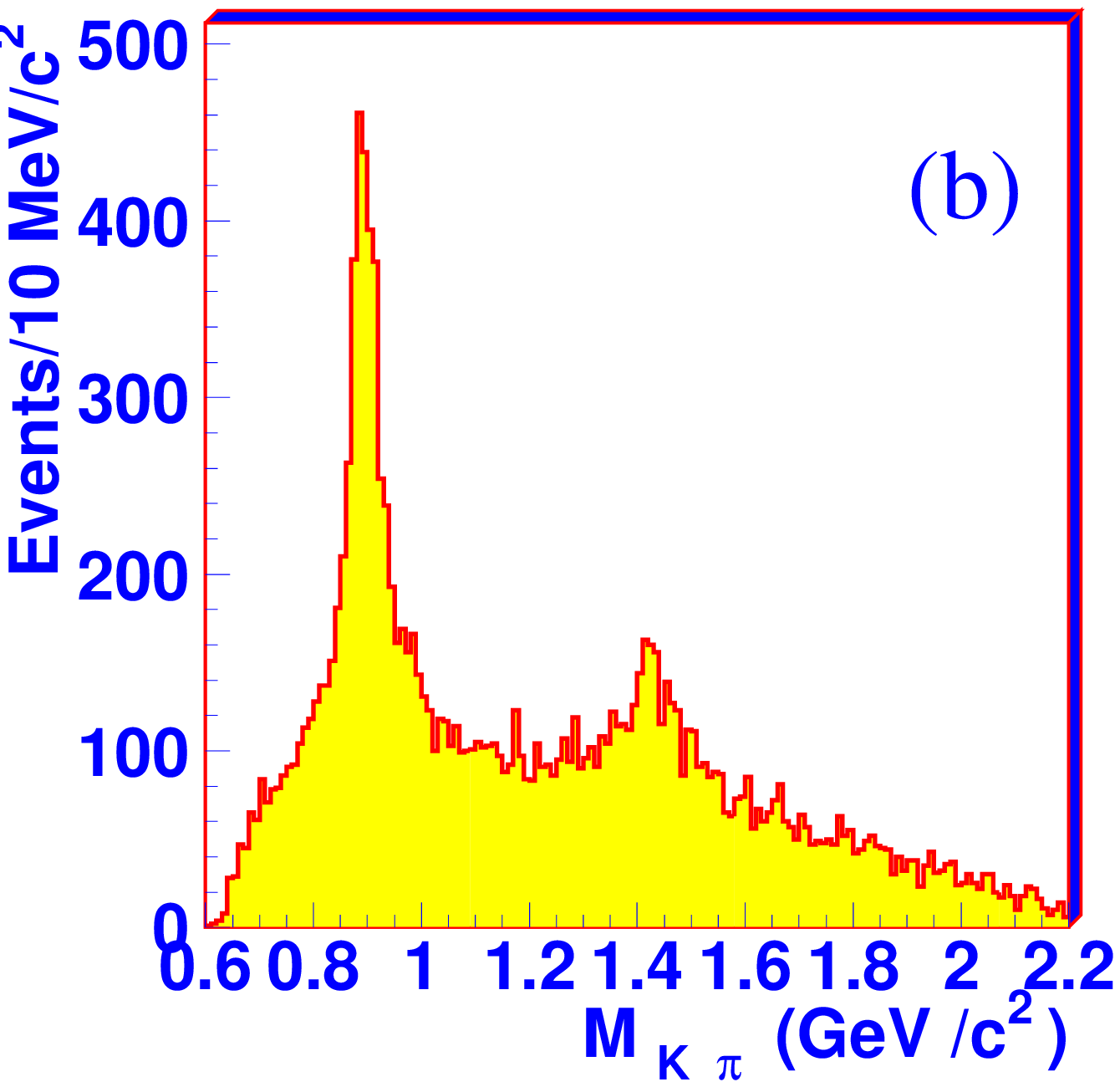,height=2.6in,width=2.6in}}}
\end{flushright}
\begin{flushleft}
{\mbox{\epsfig{file=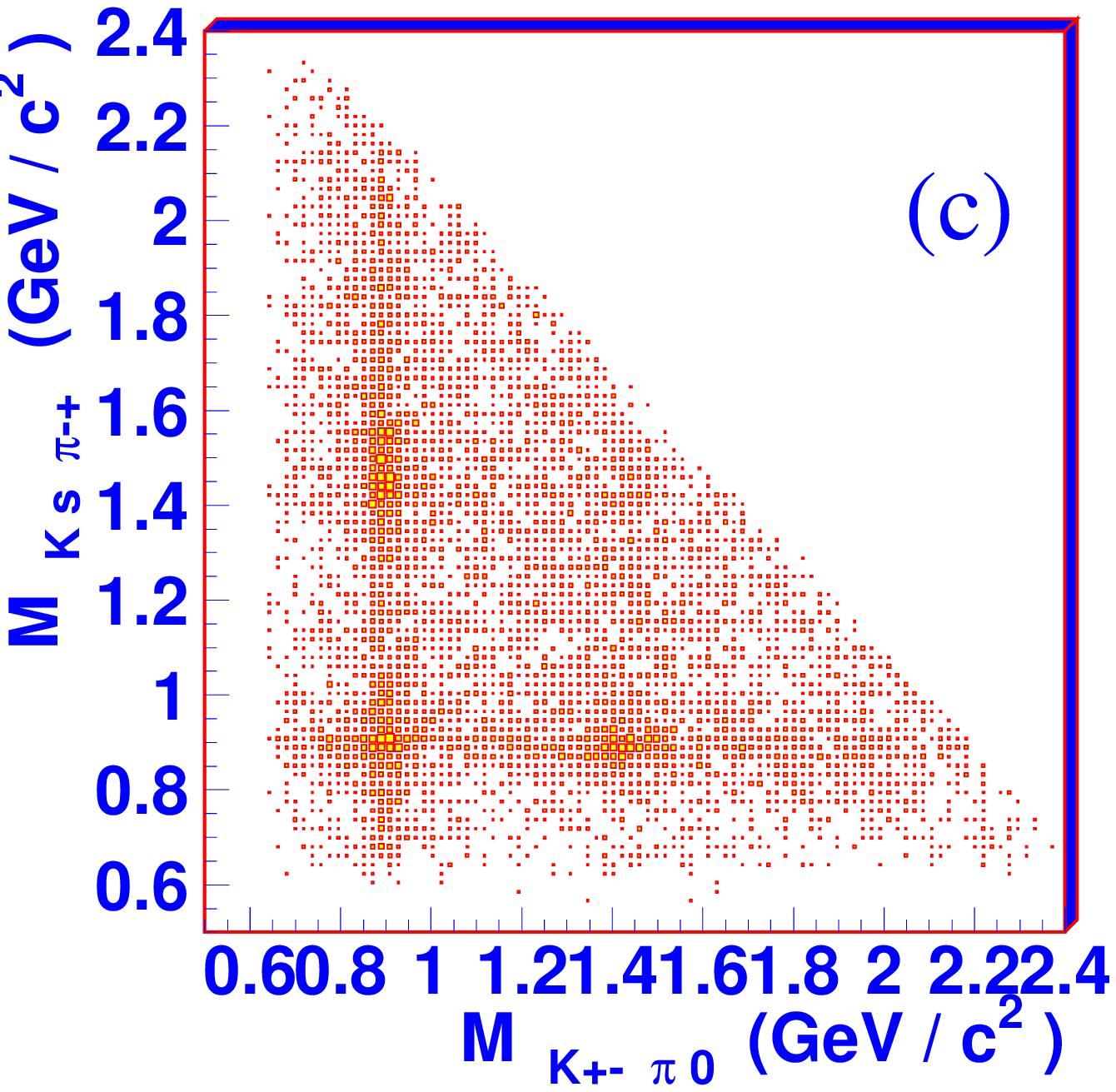,height=2.6in,width=2.6in}}}
\end{flushleft}
\vspace{-2.6in} 
\begin{flushright}
{\mbox{\epsfig{file=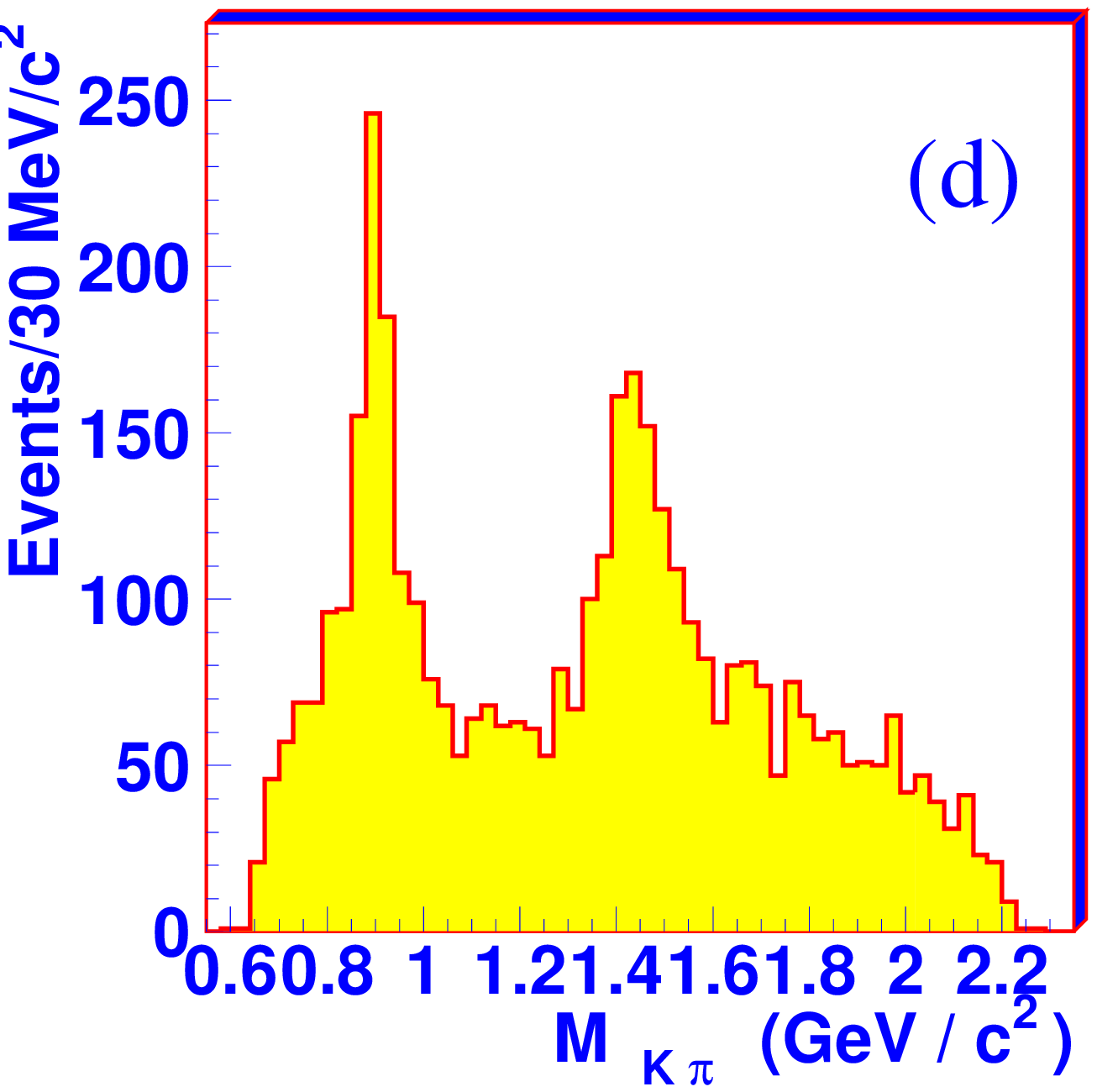,height=2.6in,width=2.6in}}}
\end{flushright}
\begin{flushleft}
{\mbox{\epsfig{file=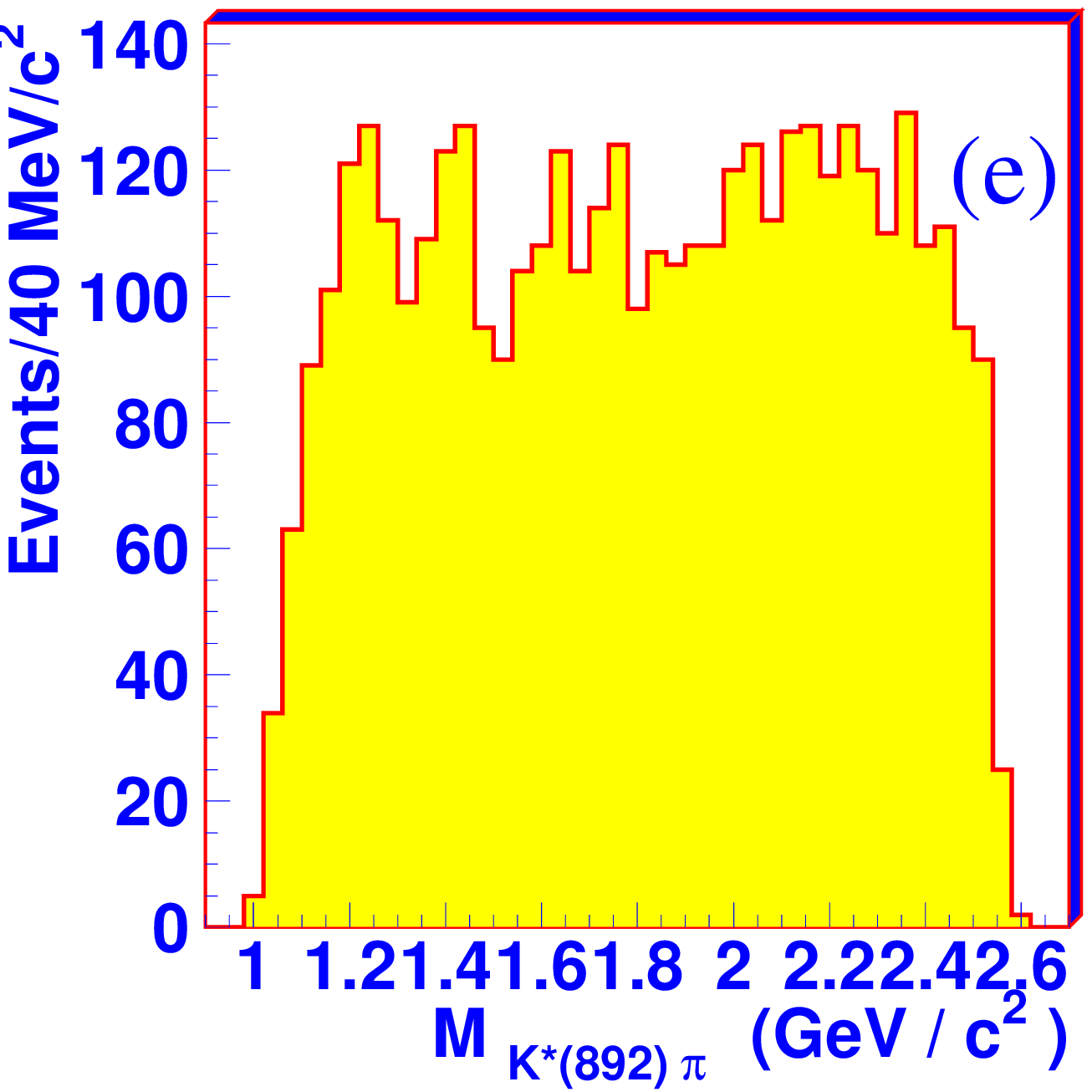,height=2.6in,width=2.6in}}}
\end{flushleft}
\vspace{-2.6in} 
\begin{flushright}
{\mbox{\epsfig{file=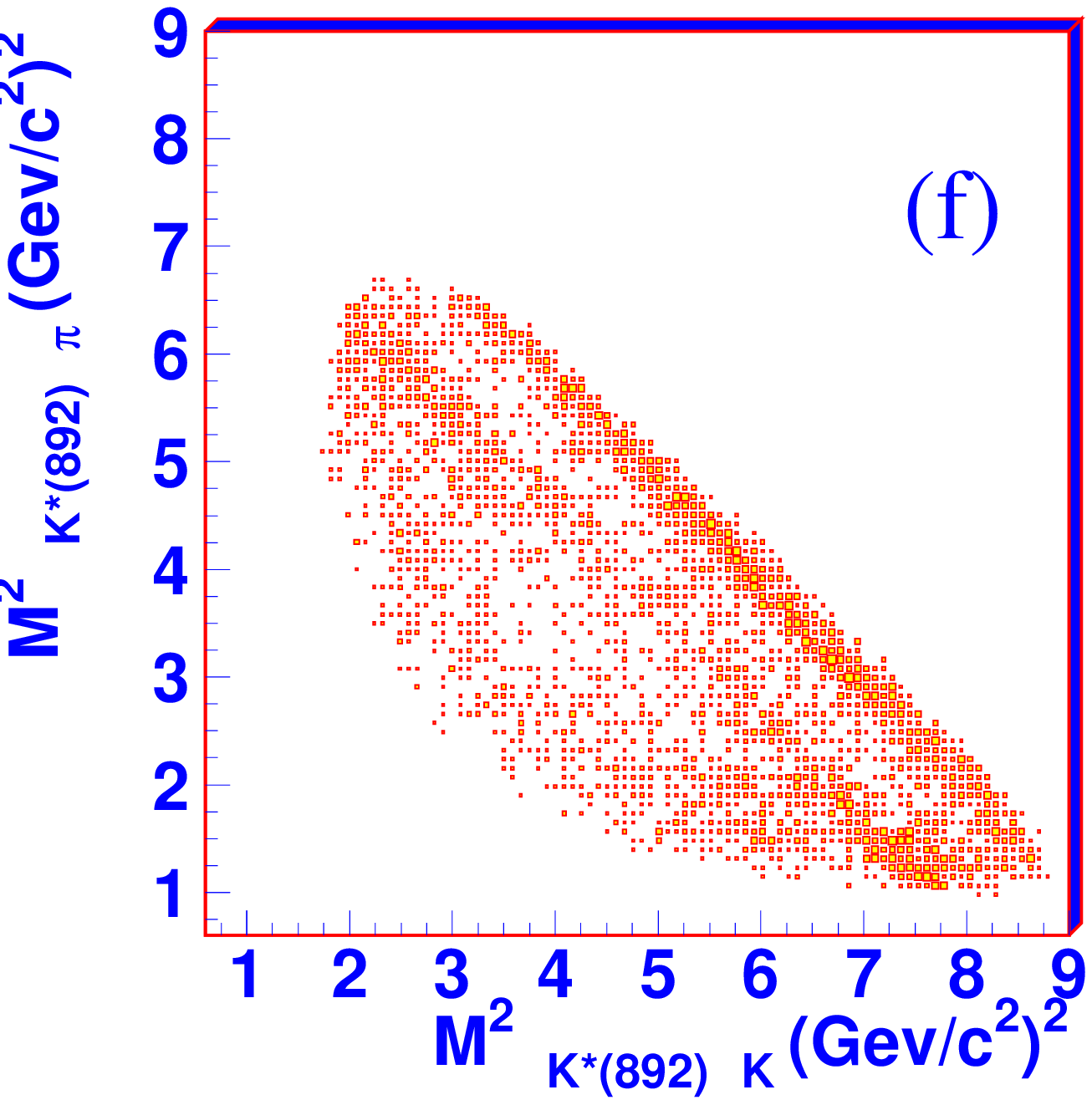,height=2.6in,width=2.6in}}}
\end{flushright}
\caption[]{  (a) $\pi^{\mp} \pi^0$ mass distribution. 
  (b) Combined $K^{\pm} \pi^0$ and $K_S \pi^{\mp}$ mass
  distribution after final cuts except the $K^*(892)$ cuts.
  (c) Scatter plot of $M_{K^{\pm}\pi^0}$ versus $M_{K_S \pi^{\mp}}$.  
  (d) Invariant mass distribution of $K^{\pm} \pi^0$ 
  and $K_S \pi^{\pm}$ recoiling against
  $K^*(892)^{\mp}$.  (e) Invariant mass distribution of $K^*(892)
  \pi$. (f) Dalitz plot.  }
\label{k01}
\end{figure}

\section{Background Studies}
Possible sources of background are studied.  First,
sideband backgrounds are studied.  The $\pi^+ \pi^-$ mass distribution
is shown in Fig.~2(a), where a clear $K_S$ signal can be seen and the
background level is quite low. The $K\pi$ spectrum from $K_S$
side-band events is shown in Fig.~2(b).  (The $K_S$ side-band is
defined by 0.02 GeV/$c^2$ $<|M_{\pi\pi}-0.497|<$ 0.04 GeV/$c^2$.)
There are no clear structures. 
The $\gamma \gamma$ spectrum is shown in Fig.~2(c), and the $K\pi$ spectrum of the 284 $\pi^0$ side-band events is shown in
Fig.~2(d).  (The $\pi^0$ sideband is defined by 0.04 GeV/$c^2$ $<
|m_{\gamma \gamma}-0.135| < $ 0.08 GeV/$c^2$.) 
The structures in Fig.~2(d) are
similar to those in the signal region and come from signal
events in the $\pi^0$ tails.  The $K^*(892)$ side-band
background is shown by the dark shaded histogram in Fig.~2(e). (The
$K^*(892)$ side-band is defined by 0.08 GeV/$c^2$ $< | M_{K \pi}
-0.892 |<$ 0.16 GeV/$c^2$.) The clear $K^*(892)$ in the $K \pi$ mass
distribution of $K^*(892)$ side-band events
mainly comes from the cross channel.  (There are two bands in the scatter
plot in Fig.~1(c). When we select one band, we will also select some events
from the other band where it crosses the first band.  These events correspond
to cross channel background.)  Fig.~2(f) shows the $K \pi$ spectrum after
side-band subtraction. The low mass enhancement and $K^*(892)$ peak
survive after side-band subtraction.

Next, we perform Monte Carlo simulation to study the main physics
background processes, including $J/\psi \to \gamma
\eta_c$ $\to \gamma K^* \bar{K}^*$ $\to \gamma K^{\pm} K_S \pi^{\mp}
\pi^0$, $J/\psi \to \gamma \eta_c$ $\to \gamma K^*(892)^{\pm} K^{\mp}
\pi^0$ $\to \gamma K^{\pm} K_S \pi^{\mp} \pi^0$, $J/\psi \to \gamma
\eta_c$ $\to \gamma K^*(892)^{\pm} K_S \pi^{\mp}$ $\to \gamma K^{\pm} K_S
\pi^{\mp} \pi^0$, $J/\psi \to \pi^0 \pi^+ \pi^- \pi^+ \pi^-$, and
$J/\psi \to \pi^0 K^+ K^- \pi^+ \pi^-$.  The selection efficiencies
are much lower than $1\%$, and the largest
number of background events contributed is about 6.
Therefore, the physics background is quite low.

\begin{figure}[htbp]
\begin{flushleft}
{\mbox{\epsfig{file=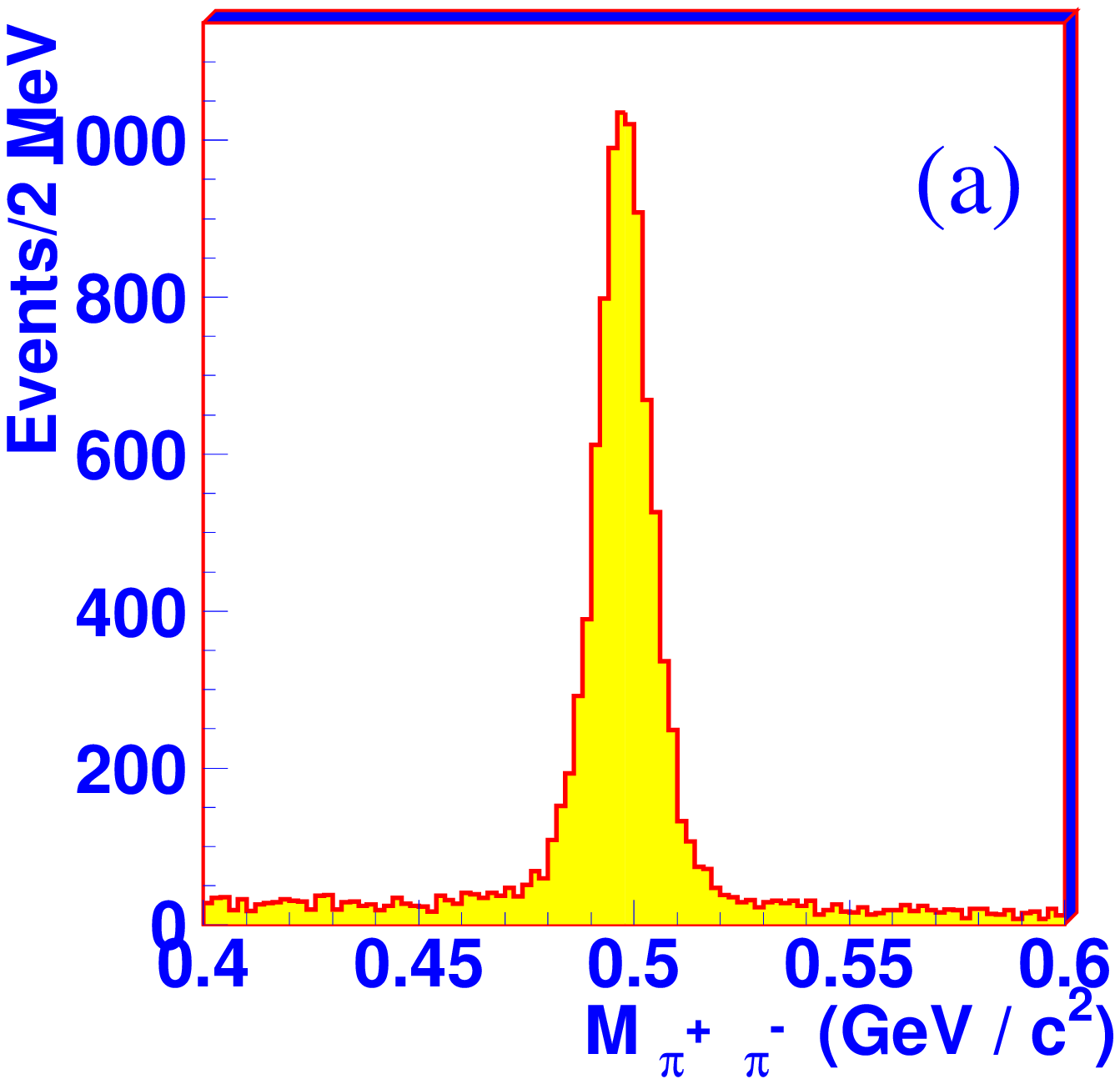,height=2.6in,width=2.6in}}}
\end{flushleft}
\vspace{-2.6in} 
\begin{flushright}
{\mbox{\epsfig{file=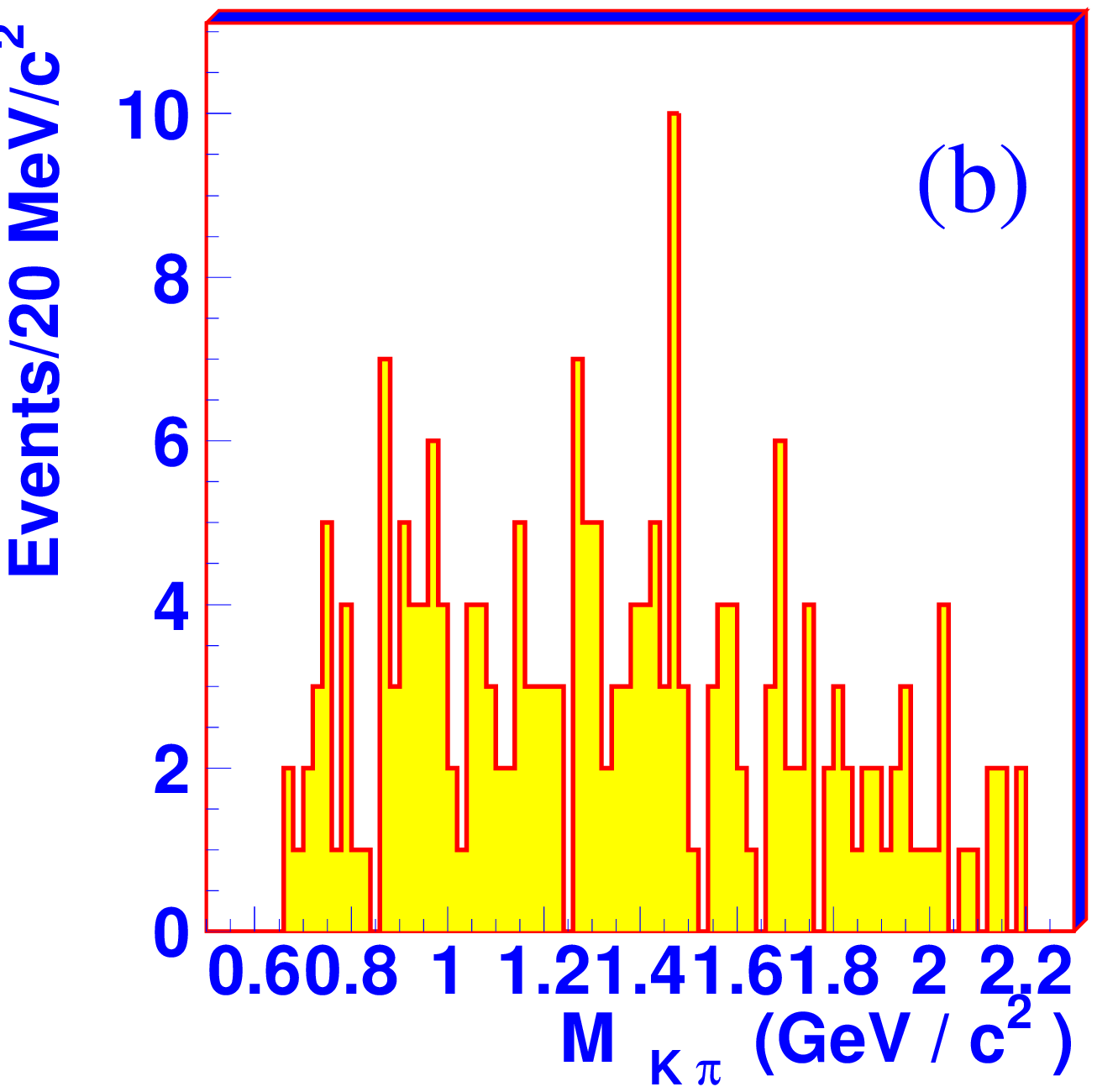,height=2.6in,width=2.6in}}}
\end{flushright}
\begin{flushleft}
{\mbox{\epsfig{file=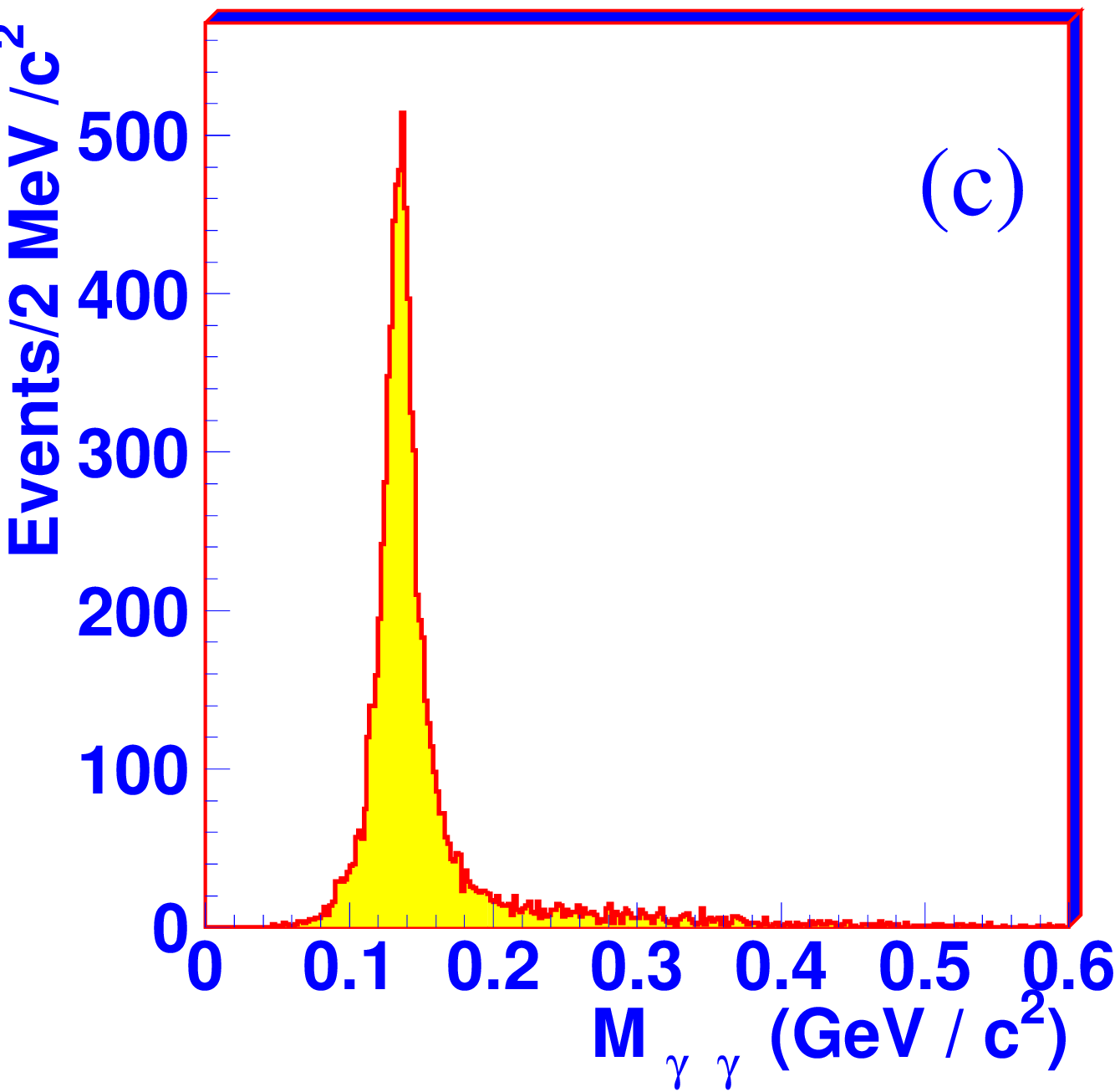,height=2.6in,width=2.6in}}}
\end{flushleft}
\vspace{-2.6in} 
\begin{flushright}
{\mbox{\epsfig{file=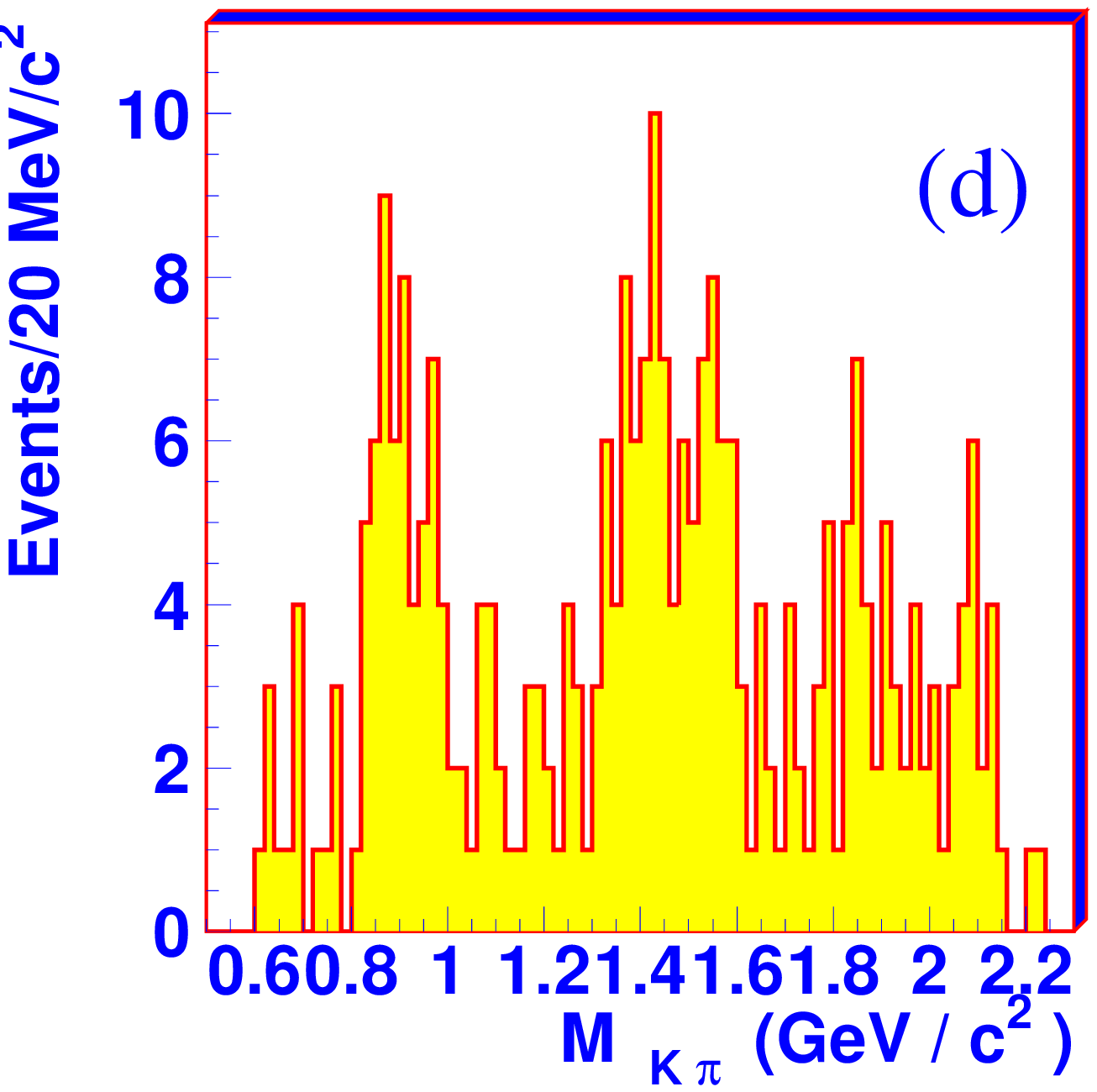,height=2.6in,width=2.6in}}}
\end{flushright}
\begin{flushleft}
{\mbox{\epsfig{file=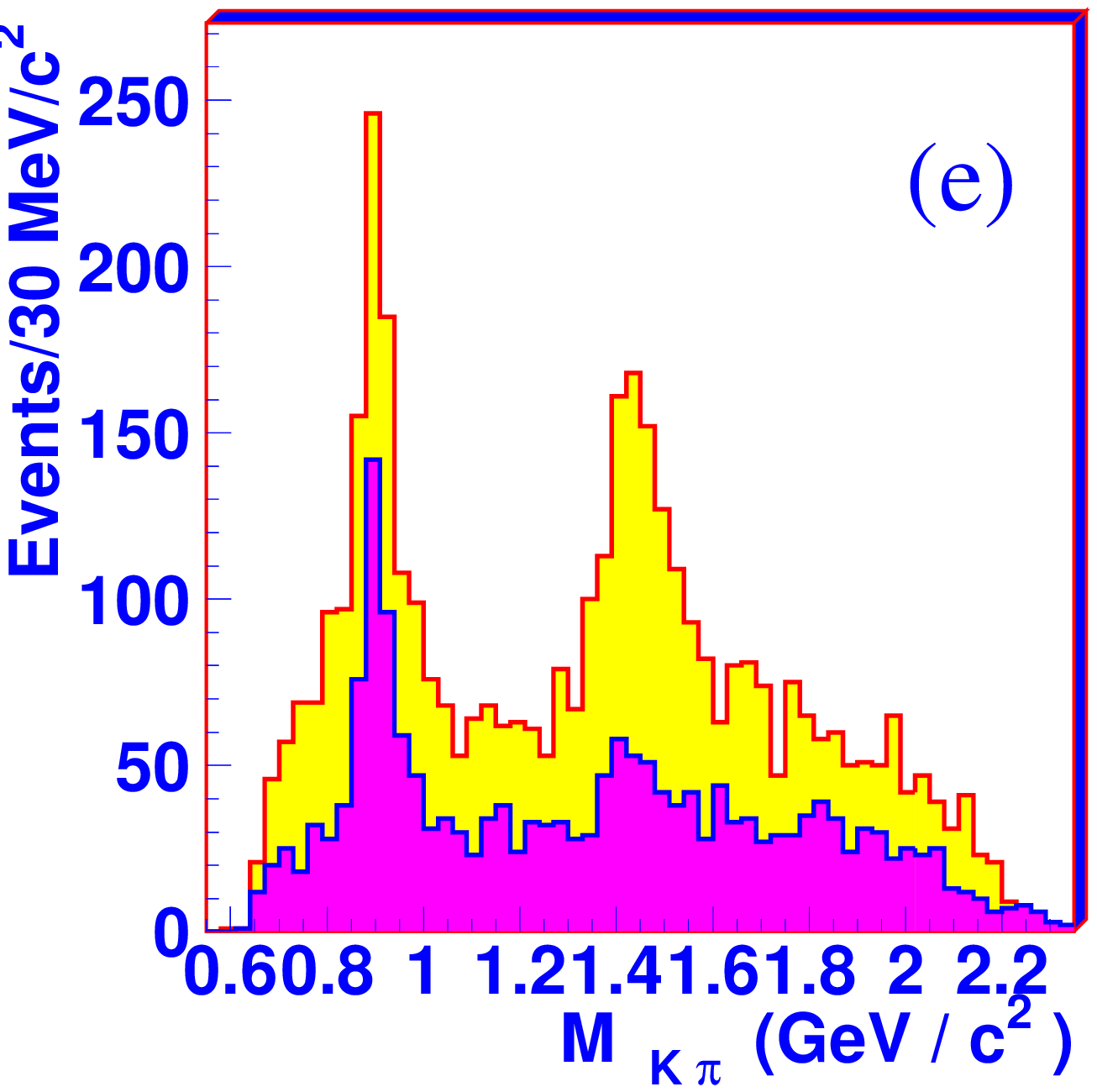,height=2.6in,width=2.6in}}}
\end{flushleft}
\vspace{-2.6in}
\begin{flushright}
{\mbox{\epsfig{file=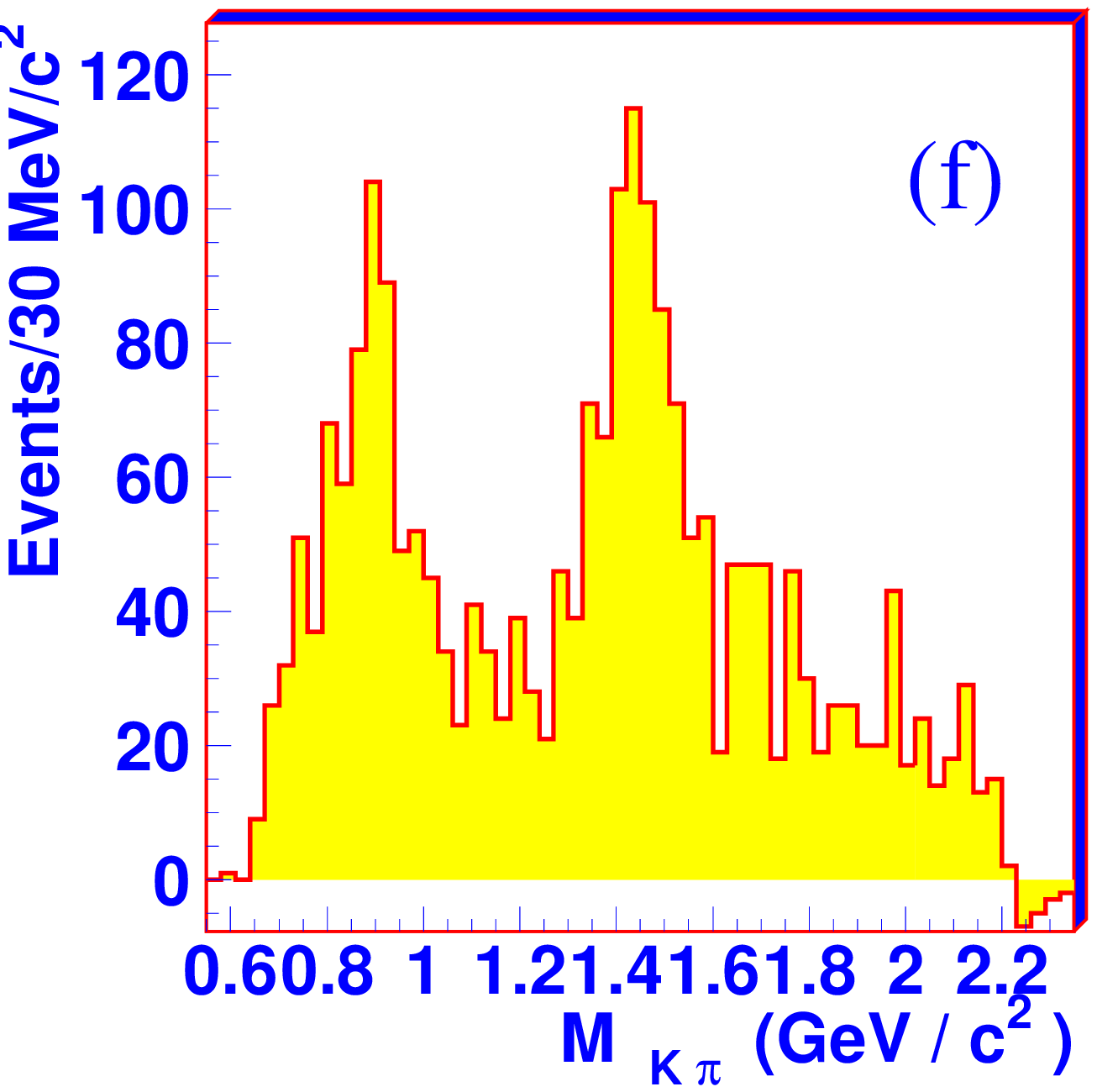,height=2.6in,width=2.6in}}}
\end{flushright}
\caption[]{(a) $m_{\pi^+\pi^-}$ mass distribution after final event selection
  except for the $K_S$ requirement.  (b) $K_S$ side-band structure.
  (c) $m_{\gamma \gamma}$ mass
  distribution after final data selection except the $\pi^0$
  requirement.  (d) $M_{K \pi}$ distribution from $\pi^0$ side-band
  events.  (e) $K \pi$ spectrum
  recoiling against $K^*(892)$.  The dark shaded histogram is from
  $K^*$ side-band events.   (f) The $K \pi$
  spectrum after side-band subtraction.  }
\label{k02}
\end{figure}

\section{Partial wave analysis}
A partial wave analysis (PWA), which is based on the
covariant helicity amplitude analysis~(\cite{jacob} - \cite{wu03}),
is performed for the charged $\kappa$.
We add the likelihoods of all four channels together, and find the
minimum of the sum.
This is the same method as
used to fit $J/\psi \to K^*(892)^0 K^+ \pi^-$\cite{bes2k}.
The main difference here is that the decay
$J/\psi \to K^*(892)^{\pm} K^*(892)^{\mp}$ is
included in the fit.

In the PWA analysis, ten resonances,  $\kappa$, $K^*_0(1430)$, $IPS$, $K_2^*(1430)$,
$K_2^*(1920)$, $K^*(1410)$ and $K^*(892)$ in the $K\pi$ spectrum,
$K_1(1270)$ and $K_1(1400)$ in the $K^*(892)\pi$ spectrum,
and $b_1(1235)$ in the $K^*(892) K $ spectrum, and two backgrounds, 
listed as the last items in Table~2, are considered in the fit. 
In Table~2, IPS (Interference Phase Space) refers to a broad
$0^{+}$ structure with a $K \pi$ invariant mass spectrum the same as
phase space,
that interferes with $\kappa$.  $K^*$ BG refers to the $K^*(892)$ background coming from the cross
channel. PS (Phase Space) refers to the background with no
interference with resonances and with a shape almost the same as that of phase space.

\begin{table}[htp]
\begin{center}
\doublerulesep 0pt
\renewcommand\arraystretch{1.1}
\begin{tabular}{|c|c|c|c|c|c|}
\hline
\hline
\hline

{\small Resonance } & {\small Spin-  } & {\small Decay}
        & {\small Mass } &{\small  Width  }
        & Sig.\\
        & {\small Parity  } & {\small Mode}
        & {\small (MeV//$c^2$) } &{\small  (MeV/$c^2$) }
        &  \\
\hline

$\kappa$ (1) &  $0^{+}$  & $K\pi$ & $810 \pm 68 ^{+15}_{-24} $
        & $536 \pm 87^{+106}_{-47}$ & $> 6 \sigma$ \\
\hline

$\kappa$ (2) &  $0^{+}$  & $K\pi$ & $884 \pm 40 ^{+11}_{-22}$
        & $478 \pm 77 ^{+71}_{-41}$ & $> 6 \sigma$  \\
\hline

$\kappa$ (3) &  $0^{+}$  & $K\pi$ &  $ 1165 \pm 58 ^{+120}_{-41}$
        & $1349 \pm 500 ^{+472}_{-176}$ & $> 6 \sigma$ \\
\hline

$K^*_0(1430)$ &  $0^{+}$  & $K\pi$ & $1400  \pm 86$
        & $325 \pm 200$ & $0.6 \sigma$  \\
\hline

IPS  &  $0^{+}$  & $K\pi$ &  --- & --- & $> 6 \sigma$ \\
\hline

$K^*_2(1430)$ &  $2^{+}$  & $K\pi$ & $1411 \pm 30$
        & $ 111 \pm 46$& $> 6 \sigma$
 \\
\hline

$K^*_2(1920)$ &  $2^{+}$  & $K\pi$ &  $2020 \pm 140$
        & $705 \pm 160 $& $> 6 \sigma$
 \\
\hline

$K^*(1410)$ &  $1^{-}$  & $K\pi$ &  $1420 \pm 14$
        & $130 \pm 28$& $> 6 \sigma$  \\
\hline

$K^*(892)$ &  $1^{-}$  & $K\pi$ &  $ 896 \pm 8$
        & $57 \pm 12$ &$> 6 \sigma$  \\
\hline

$K_1(1270)$ &  $1^{+}$  & $K^*(892) \pi$ & $1254 \pm 14$
        & $60 \pm 28$& $> 6 \sigma$  \\
\hline

$K_1(1400)$ &  $1^{+}$  & $K^*(892) \pi$ &  $1390 \pm 30 $
        & $146 \pm 44$& $> 6 \sigma$ \\
\hline

$b_1(1235)$ &  $1^{+}$  & $K^*(892) K$ &  $1230 \pm 52$
        & $142 \pm 38$  &$ 4.5 \sigma$ \\
\hline

$K^*(892)$ BG &  $1^{-}$  & $K\pi$ &  --- & --- & $> 6 \sigma$ \\
\hline

PS  BG &  ---  & --- & ---  & --- & --- \\
\hline

\hline
\end {tabular}
\vspace{0.1in}
\caption {Resonances included in the fit of this channel.
Masses and widths of various resonances are determined by
mass and width scans. $\kappa$ (1), (2) and (3) are results
given by fits using Breit-Wigner functions (1), (2) and (3)
to fit $\kappa$ respectively. IPS refers to the broad $0^+$ structure which interferes with $\kappa$.
$K^*$ BG refers to the $K^*(892)$ background coming from the cross
channel. PS BG refers to the background with
no interference with resonances.}
\end{center}
\end{table}

Three different parameterizations
are used to fit the $\kappa$. They are
\be\label{b1}
BW_{\kappa} = \frac{1}{m_{\kappa}^2 -s - i m_{\kappa} \Gamma_{\kappa}},
~~~~~ \Gamma_{\kappa} = {\rm constant} ,
\ee
\be\label{b2}
BW_{\kappa} = \frac{1}{m_{\kappa}^2 -s - i \sqrt{s} \Gamma_{\kappa}(s)},
~~~~~ \Gamma_{\kappa}(s) =
        \frac{g_{\kappa}^2 \cdot k_{\kappa}}{8 \pi s} ,
\ee
\be\label{b3}
BW_{\kappa} = \frac{1}{m_{\kappa}^2 -s - i \sqrt{s} \Gamma_{\kappa}(s)},
~~~~~ \Gamma_{\kappa}(s) = \alpha \cdot k_{\kappa} ,
\ee
where $k_{\kappa}$ is the magnitude of the $K$ momentum
in the $K \pi$, or the $\kappa$,
center of mass system~\cite{hanqing2}, and $\alpha$ is
a constant which will be determined by the fit.
Parameters in the Breit-Wigner function are determined
by mass and width scans. The minima of the scan curves give 
the central values of mass and width parameters.
From these, the corresponding Breit-Wigner pole positions can be
directly calculated from equation (1), (2) and (3).
Our final results are listed in Table~3, where
the first errors are statistical, and
the second are systematic.
The mass and width parameters
obtained by different parameterizations are quite different,
but their poles are almost the same, which
is quite similar to what was found in the study of the neutral
$\kappa$. The results for the neutral
$\kappa$~\cite{bes2k} are shown in Table~4 and are consistent with the
charged $\kappa$. 

\begin{table}[htp]
\begin{center}
\doublerulesep 0pt
\renewcommand\arraystretch{1.1}
\begin{tabular}{|c|c|c|c|}
\hline

 & BW (1)  &  BW (2)  & BW (3)  \\
\hline

Mass (MeV/$c^2$) & $810 \pm 68 ^{+15}_{-24} $
        &  $884 \pm 40 ^{+11}_{-22} $ &
        $1165 \pm 58 ^{+120}_{-41} $ \\
\hline

Width (MeV/$c^2$) & $ 536 \pm 87^{+106}_{-47} $
        & $478 \pm 77 ^{+71}_{-41} $
        & $1349 \pm 500 ^{+472}_{-176} $  \\
\hline

pole (MeV/$c^2$) & $(849 \pm 77 ^{+18}_{-14} )$
        & $(849 \pm 51 ^{+14}_{-28} $)
        & $(839 \pm 145 ^{+24}_{-7})$  \\

        & $-i(256 \pm 40 ^{+46}_{-22} )$
        & $-i (288 \pm 101 ^{+64}_{-30} )$
        & $-i(297 \pm 51 ^{+50}_{-18})$ \\

\hline
\end {tabular}
\vspace{0.1in}
\caption {Masses, widths and pole positions of the charged
        $\kappa$. In the table, the first errors
        are statistical, and the second are systematic.
BW (1) means equation (1) is
        used to fit the $\kappa$. BW (2) and BW (3)
        have similar meanings. }
\end{center}
\end{table}

\begin{table}[htp]
\begin{center}
\doublerulesep 0pt
\renewcommand\arraystretch{1.1}
\begin{tabular}{|c|c|c|c|}
\hline

 & BW (1) &  BW (2) & BW (3)  \\
\hline

Mass (MeV/$c^2$) & $745 \pm 26^{+14}_{-91}$
        & $ 874 \pm 25 ^{+12}_{-55}$  &  $1140 \pm 39 ^{+47}_{-80}$ \\
\hline

Width (MeV/$c^2$) &  $622 \pm 77 ^{+61}_{-78} $
        & $518 \pm 65 ^{+27}_{-87}$  & $1370 \pm 156 ^{+406}_{-148}$  \\
\hline

pole (MeV/$c^2$) & $(799 \pm 37 ^{+16}_{-90})$
        & $(836 \pm 38 ^{+18}_{-87})$
        & $(811 \pm 74 ^{+17}_{-83})$ \\

 & $-i(290 \pm 33 ^{+25}_{-38})$
 & $-i(329 \pm 66 ^{+28}_{-46})$
 & $-i(285 \pm 20 ^{+18}_{-42})$  \\
\hline
\end {tabular}
\vspace{0.1in}
\caption {Masses, widths and pole positions of the neutral
        $\kappa$~\cite{bes2k}. BW (1) means equation (1) is
        used to fit the $\kappa$. BW (2) and BW (3)
        have similar meaning.}
\end{center}
\end{table}

Our final results correspond to
the solution with the minimum least likelihood. Differences among
solutions with similar likelihood values are included as
systematic uncertainties.  Also included are
the effect of removing $K_0(1430)$,
IPS, $b_1(1235)$ and $K^*(892) \bar{K}^*(892)$,  the result of a
fit with the $K^*(892)$ background level floating, and the result from
a fit using direct side-band subtraction.

The masses and widths of all resonances obtained
by mass and width scans are shown in Table~2.
In the fit, the contribution from $K^*_0(1430)$
is small; its statistical significance is only 0.6$\sigma$.
Because it is expected in this channel, it is included in the
final solution. 
The $K \pi $ mass distribution is shown in Fig.~3(a), where points with
error bars are data, and the light shaded histogram is the final fit.
In the figure,
the dark shaded histogram shows the contribution of the charged
$\kappa$.    Fig.~3(b) shows the fit for the
$K^*(892) \pi $ spectrum, and Fig.~4 shows the
angular distributions. 

\begin{figure}[htbp]
\begin{flushleft}
{\mbox{\epsfig{file=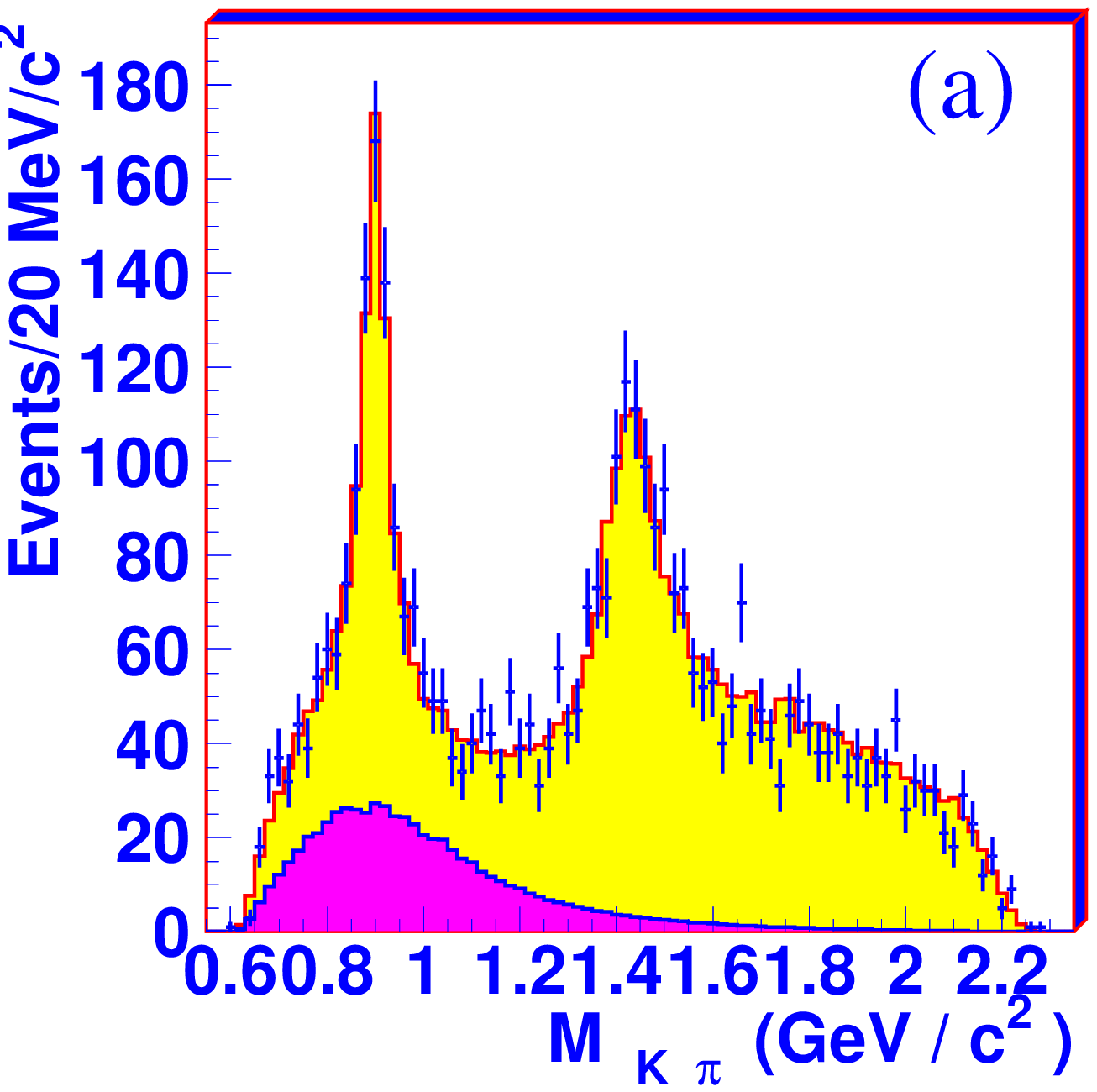,height=2.6in,width=2.6in}}}
\end{flushleft}
\vspace{-2.6in}
\begin{flushright}
{\mbox{\epsfig{file=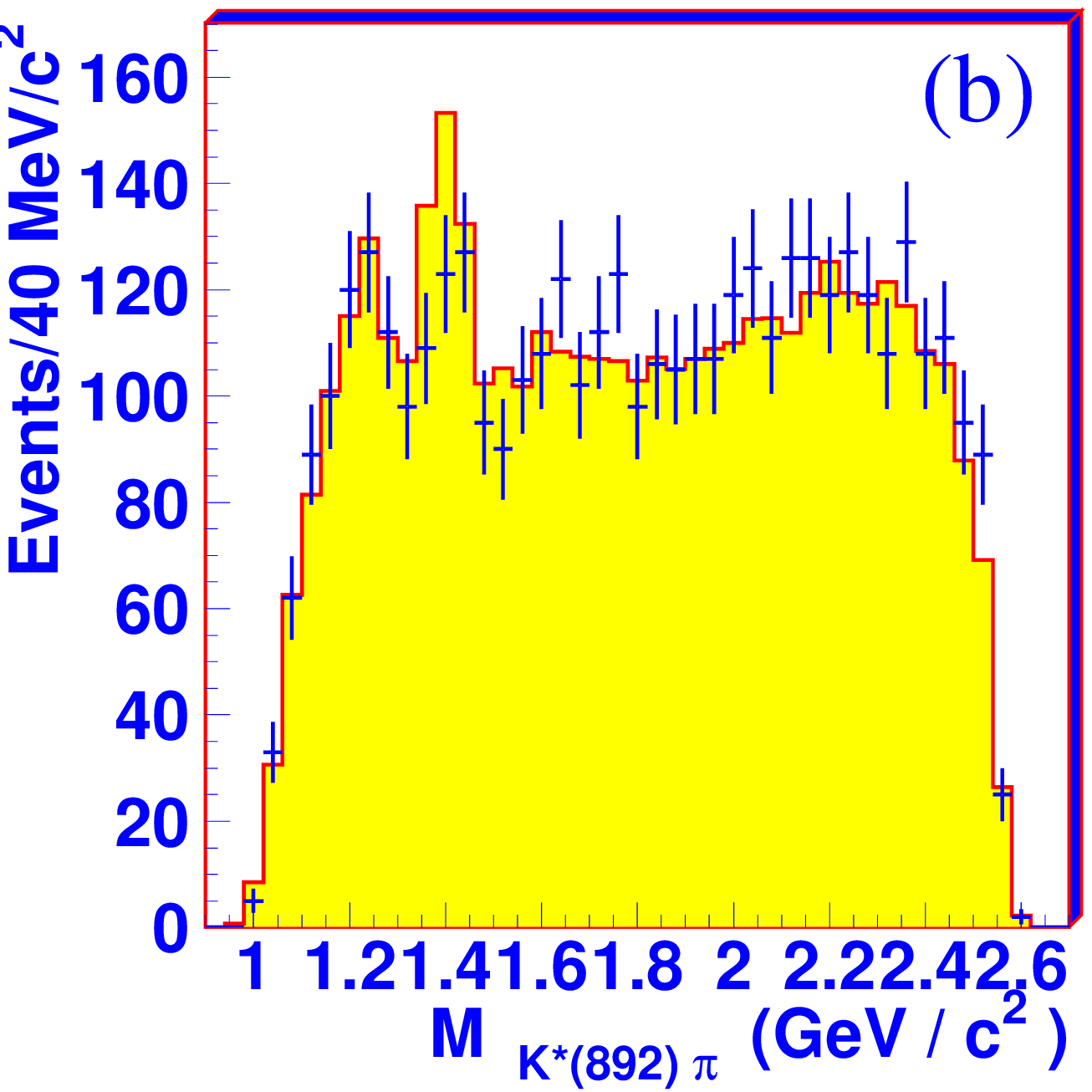,height=2.6in,width=2.6in}}}
\end{flushright}
\caption[]{(a) Final fit results for the $K \pi $ spectrum. Points
  with error bars
  are data,
        the light shaded histogram is the final global fit, and the dark
        shaded histogram is the contribution of the $\kappa$.
        (b) Final fit of $K^*(892) \pi $ spectrum.
        Dots with error bars are data, and
        the histogram is the final global fit. There are two peaks
        in the lower mass region. The lower one is fit by the $K_1(1270)$,
        and the higher one is by the $K_1(1400)$. }
\label{k05}
\end{figure}

\begin{figure}[htbp]
\begin{flushleft}
{\mbox{\epsfig{file=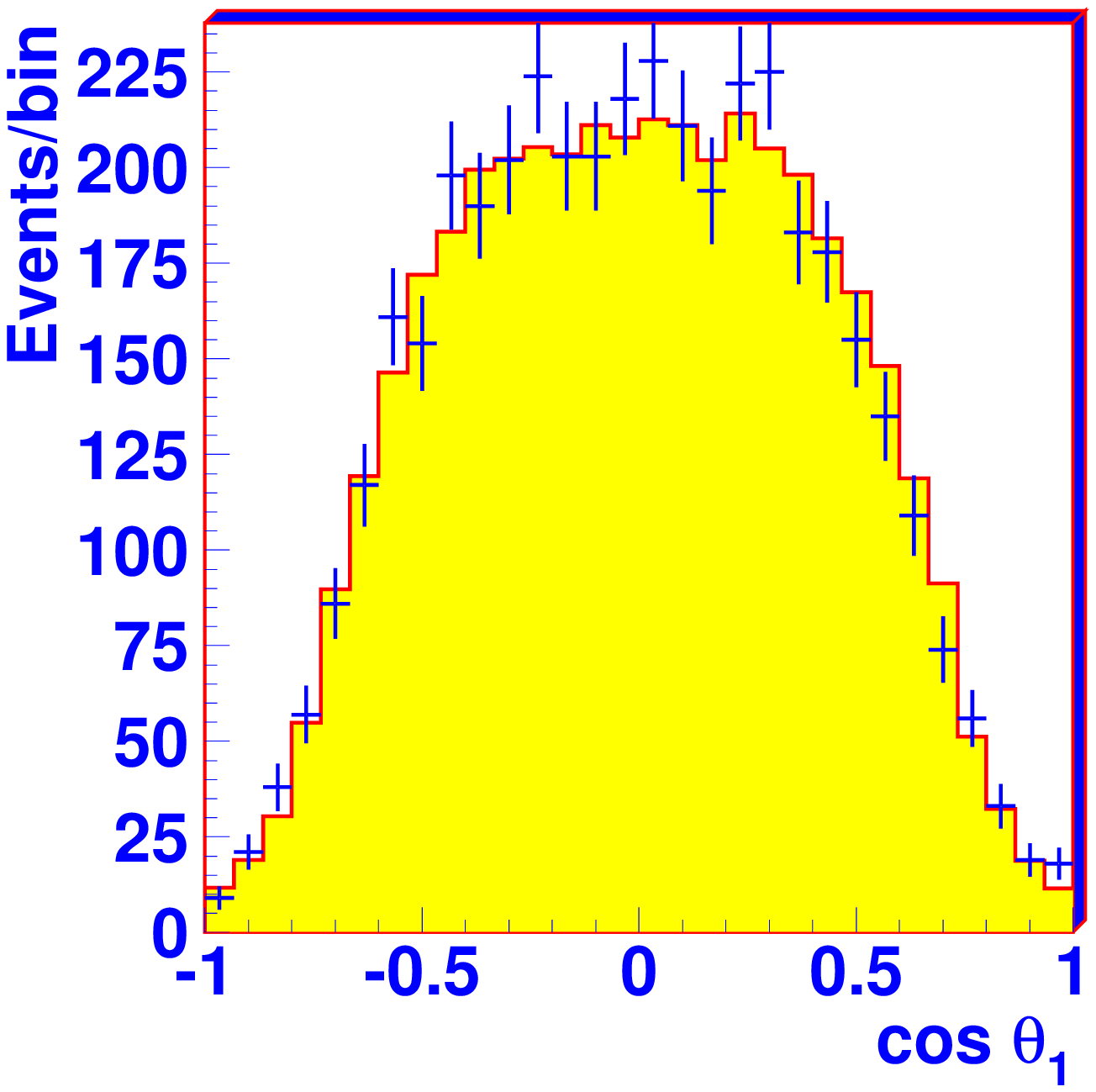,height=2.6in,width=2.6in}}}
\end{flushleft}
\vspace{-2.6in} 
\begin{flushright}
{\mbox{\epsfig{file=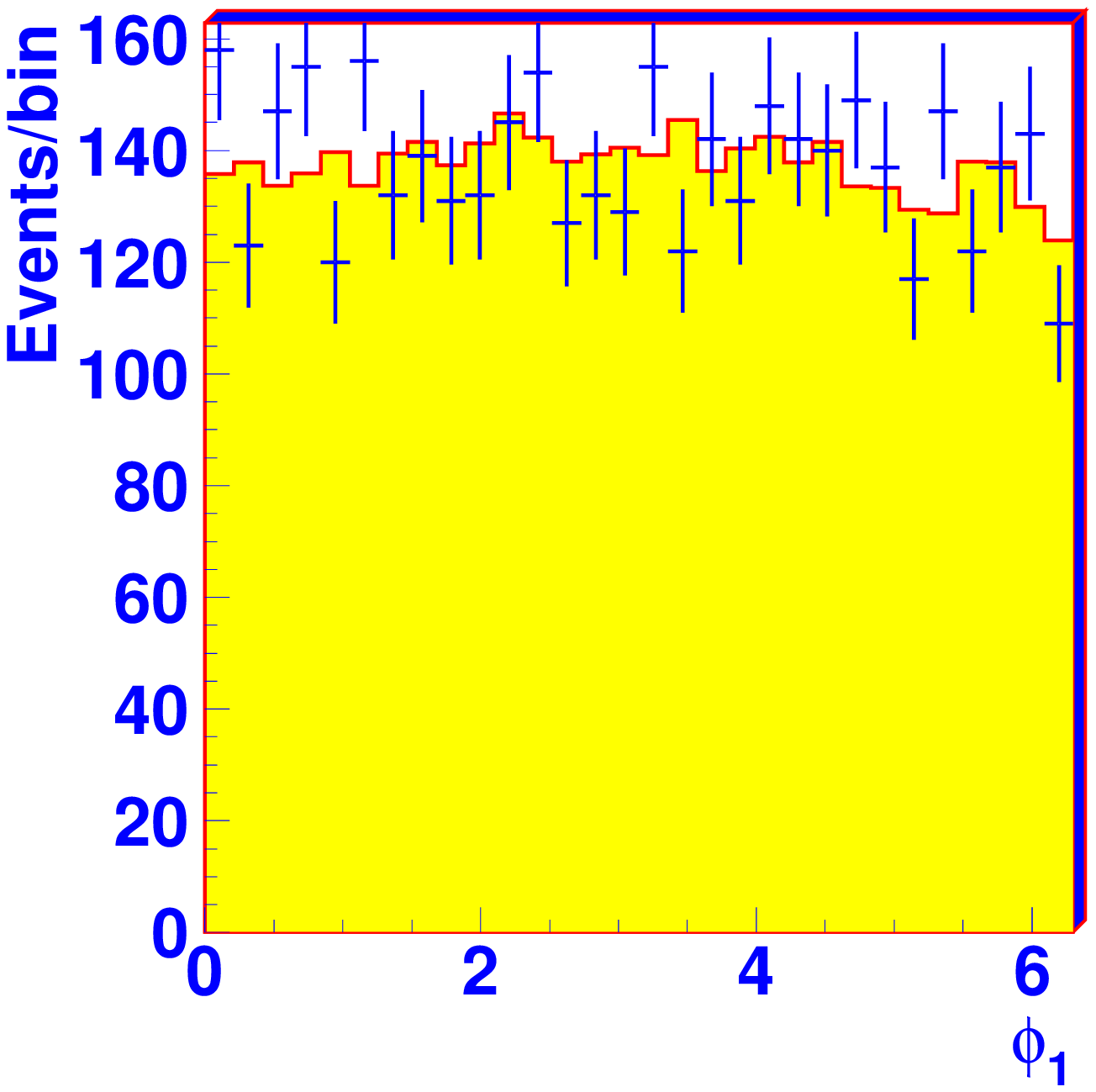,height=2.6in,width=2.6in}}}
\end{flushright}
\begin{flushleft}
{\mbox{\epsfig{file=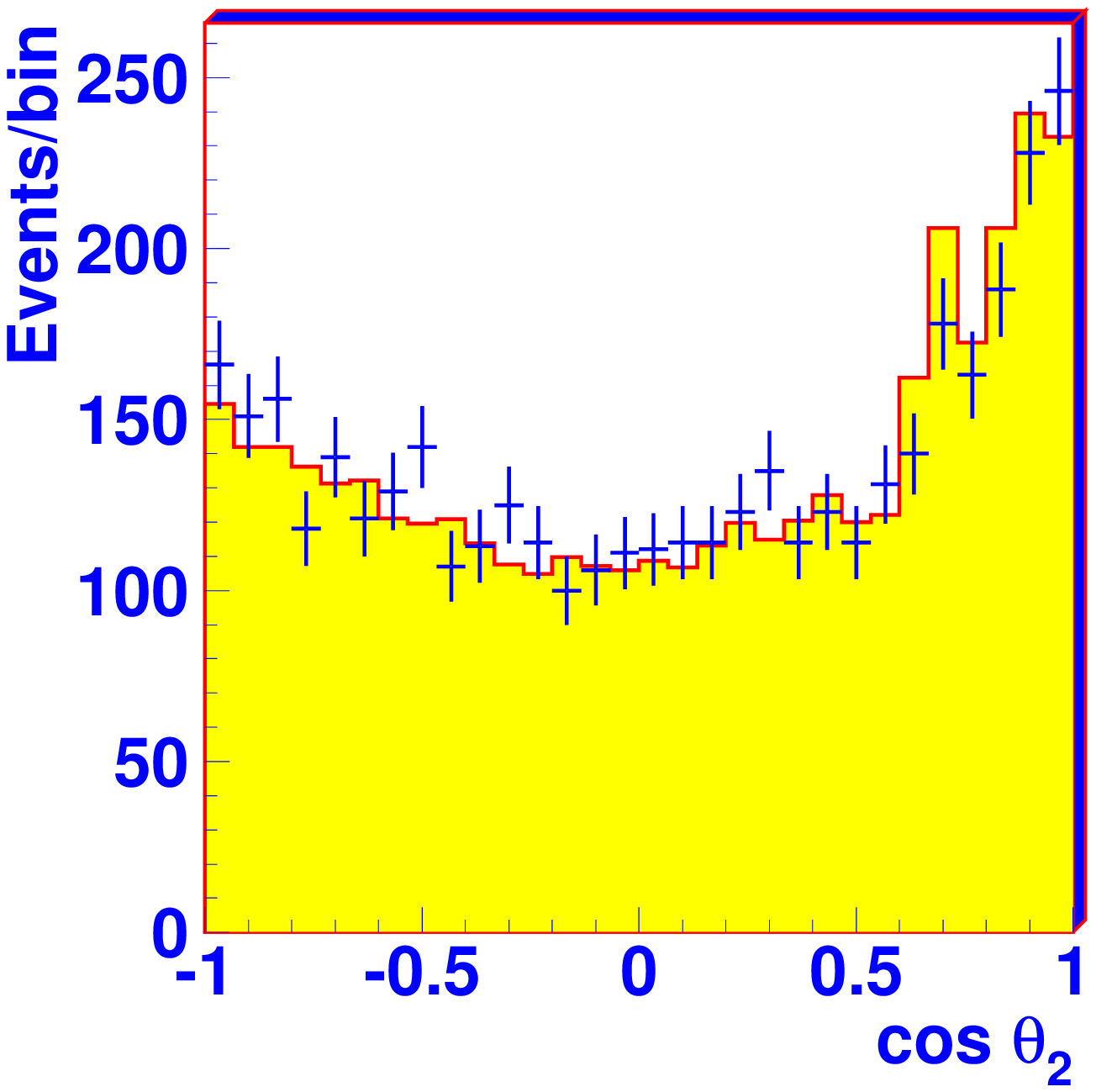,height=2.6in,width=2.6in}}}
\end{flushleft}
\vspace{-2.6in} 
\begin{flushright}
{\mbox{\epsfig{file=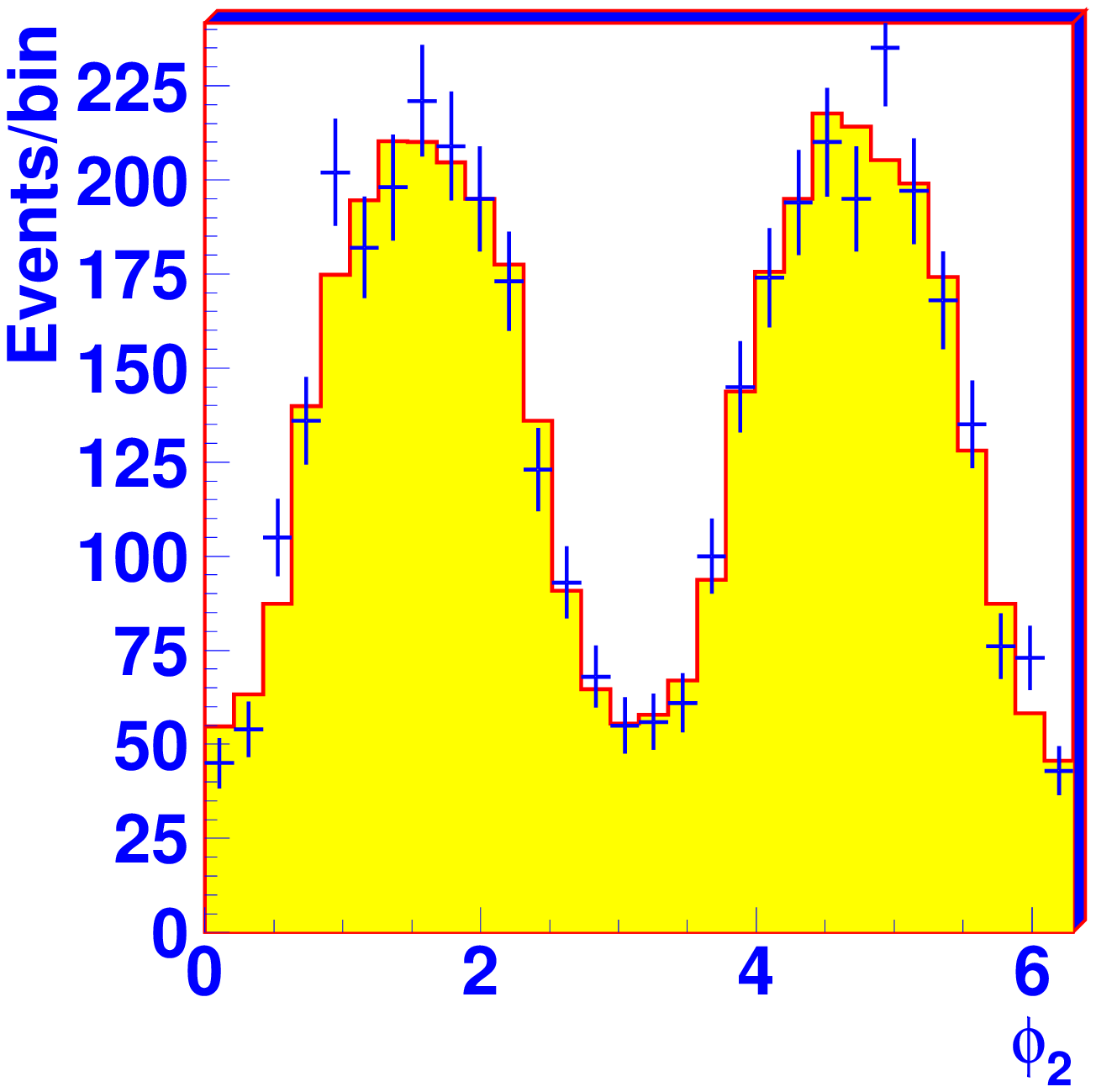,height=2.6in,width=2.6in}}}
\end{flushright}
\caption[]{ Final fit result for the angular distributions.
  Dots with error bars are data, and the histogram is the final global
  fit.  The upper left figure shows the final fit of the $\theta_1$
  distribution, the upper right shows the final fit of the $\phi_1$
  distribution, the lower left shows the final fit of the $\theta_2$
  distribution, the lower right shows the final fit of the $\phi_2$
  distribution.  $\theta_1$ and $\phi_1$ are the polar angle and
  azimuthal angle of the $K \pi$ system in the $J/\psi$ center of mass
  system.  $\theta_2$ and $\phi_2$ are the polar angle and azimuthal
  angle of the $K$ meson in the $K \pi$ center of mass system.  }
\label{k06}
\end{figure}

\section{Branching ratio measurements}
The decay
$J/\psi \to K^*(892)^{\pm} \kappa^{\mp}$
contributes 655 events. Monte Carlo simulation
of
$J/\psi \to K^*(892)^{\pm} \kappa^{\mp}
\to K^{\pm} K_S \pi^{\mp} \pi^0$
determines an efficiency of 2.33\%,
and the branching ratio of
$J/\psi \to K^*(892)^+ \kappa^-$ or
$J/\psi \to K^*(892)^- \kappa^+$
is
\be
BR = \frac{655/4}{2.33\% \times 5.8\times 10^7 \times 1/9 }
= ( 1.09 \pm 0.18 ^{+0.94}_{-0.54} ) \times 10^{-3},
\ee
where, the first error is statistical, the second error
is systematic, the factor $\frac{1}{4}$ is because the events
are the sum of four channels, and $\frac{1}{9}$ is
the isospin factor.
In the previous study of the decay
$J/\psi \to \bar{K}^*(892)^0 K^+ \pi^-$,
the corresponding branching ratio for the neutral $\kappa$
\footnote{The branching ratio of the neutral $\kappa$ was not reported in
        Ref.~\cite{bes2k}. In the study of the neutral $\kappa$, the number
        of $\kappa$ events was in the range 1891 - 3516, and
        the selection efficiency was 14.2\%. Its branching ratio is
        $\frac{1891 - 3516}{14.2\% \times 5.8 \times 10^7 \times 4/9} =$
        $(0.52 - 0.97) \times 10^{-3}$}
is $(0.52 - 0.97) \times 10^{-3}$.
The two results are consistent
with isospin symmetry.
The systematic error includes
uncertainties from multi-solutions, from different background fit
methods, and from removing some components
from the fit ($K^*_0(1430)$, IPS, $b_1(1235)$,
and $K^*(892)^{\pm} K^*(892)^{\mp}$).

Also interesting is the existence
of the $J/\psi$ electromagnetic decay
$J/\psi \to K^*(892)^{\pm} K^*(892)^{\mp}$.
The peak at 892 MeV/$c^2$ in the $K \pi$ invariant
mass spectrum (see Fig.~1(d)) comes from both
cross channel background and
from $J/\psi \to K^*(892)^{\pm} K^*(892)^{\mp}$.
The number of events from the cross channel
can be calculated approximately from the $K^*(892)$
side-band structure, shown in
Fig.~2(e), where the narrow peak at
892 MeV/$c^2$ is clear. In the final fit, the decay
$J/\psi \to K^*(892)^{\pm} K^*(892)^{\mp}$
contributes 323 events. The Monte Carlo simulation
of the decay
$J/\psi \to K^*(892)^{\pm} K^*(892)^{\mp}
\to K^{\pm} K_S \pi^{\mp} \pi^0$
yields an efficiency of 1.25\%,
and the branching ratio of
$J/\psi \to K^*(892)^+ K^*(892)^-$
is
\be
BR = \frac{323/2}{1.25\% \times 5.8\times 10^7 \times 1/9 \times 2}
= ( 1.00 \pm 0.19 ^{+0.11}_{-0.32} ) \times 10^{-3},
\ee
where, the first error is statistical, the second error
is systematic, the factor $\frac{1}{2}$ is because events are counted
 twice, $\frac{1}{9}$ is the isospin
factor, and 2 is because the data
comes from two decay channels:
$J/\psi \to K^*(892)^+ K^*(892)^-
\to (K^+ \pi^0) (K_S \pi^-) $
and
$J/\psi \to K^*(892)^- K^*(892)^+
\to (K^- \pi^0) (K_S \pi^+).$
The systematic error includes
uncertainties from multi-solutions, from different background fit
methods, and from removing some components
from fit($K^*_0(1430)$, IPS, and $b_1(1235)$).

\section{Summary}
In conclusion, the charged $\kappa$ is observed
and studied in the  decay
$J/\psi \to K^*(892)^{\pm} \kappa^{\mp}
\to K^{\pm} K_S \pi^{\mp} \pi^0$.
The low mass enhancement in the $K \pi$ spectrum cannot be
fit well unless a charged $\kappa$ is added into the solution.
If we use a Breit-Wigner
function of constant width to parameterize the $\kappa$, its
pole locates at
$(849 \pm 77 ^{+18}_{-14}) -i (256 \pm 40 ^{+46}_{-22})$ MeV/$c^2$.
In our analysis, three  different $\kappa$ parameterizations
are tried in the fit, and final results are shown in Table~3 and
are consistent with those of the neutral $\kappa$ and are also
in good agreement with those obtained in the analysis of
$K \pi$ scattering phase shifts.
Also, the decay $J/\psi \to K^*(892)^{\pm}
K^*(892)^{\mp}$ is observed for the first time with
the branching ratio
$(1.00 \pm 0.19 ^{+0.11}_{-0.32}) \times 10^{-3}$.
The corresponding decay mode is not observed
in $J/\psi \to \bar{K}^*(892)^0 K^+ \pi^-$.
The decays
$J/\psi \to K^*(892)^{\pm} K^*(892)^{\mp} $
can be produced through $J/\psi$ electromagnetic
decays, while $J/\psi \to K^*(892)^0 \bar{K}^*(892)^0 $
can only be produced through $J/\psi$ hadronic decays, which
would be $SU(3)$ symmetry breaking decays and are suppressed.
\vspace{3mm}

{\bf Acknowledgments}
The BES Collaboration thanks the staff of BEPC and computing
center for their hard efforts. This work is supported in part by
the National Natural Science Foundation of China under contract
Nos. 10491300, 10225524, 10225525, 10425523, 10625524, 10521003, 10821063,
10825524, the Chinese Academy of Sciences under contract No. KJ 95T-03, the
100 Talents Program of CAS under Contract Nos. U-11, U-24, U-25,
and the Knowledge Innovation Project of CAS under contract Nos.
U-602, U-34 (IHEP), the National Natural Science Foundation of
China under contract Nos. 10775077, 10225522 (Tsinghua University), and the
Department of Energy under Contract No. DE-FG02-04ER41291 (U.
Hawaii).


\begin{thebibliography}{99}

\bibitem{bes1s} Ning Wu (BES collaboration), "BES R measurements
        and $J/\psi$ Decays", Proceedings of the XXXVIth
        Rencontres de Moriond, Les Arcs, France,
        March 17 -- 24, 2001, Ed. J. Tran Thanh Van.
        2001 QCD and High Energy
        Hadronic Interactions,  p.3-6.

\bibitem{d1} J.Z.Bai, BES Collaboration, hep-ex/0304001.

\bibitem{bes1s2} J.Z.Bai, et al., BES collaboration, High Energy
        Phys. Nucl. Phys. 28 (2004) 215.

\bibitem{bes2s} M. Ablikim, et al., BES collaboration,
        Phys. Lett. {\bf B 598} (2004) 149.

\bibitem{bes2sp} M. Ablikim, et al., BES collaboration,
        Phys. Lett. {\bf B 645} (2007) 19.

\bibitem{bes2k} M. Ablikim, et al., BES collaboration,
        Phys. Lett. {\bf B 633} (2006) 681.

\bibitem{s0} E.M. Aitala, et al., Fermilab E791 Collaboration,
        Phys. Rev. Lett. 86 (2001)770.

\bibitem{s0k} E.M. Aitala, et al., Fermilab E791 Collaboration,
        Phys. Rev. Lett. 89 (2002)121801.



\bibitem{s1} D. Alde, et al., Phys. Lett {\bf B 397}
        (1997) 350.

\bibitem{s2} T. Ishida, et al., in: Proceedings of International
        Conference Hadron'95, World Scientific, Manchester, UK, 1995.

\bibitem{s3} V.E. Markushin, M.P. Locher, Frascati Phys.
        ser. 15 (1999) 229.

\bibitem{s4} Z.Xiao, H.Q.zheng, Nucl. Phys. A 695 (2001) 273.



\bibitem{d2} J.M. Link, et al., FNAL FOCUS Collaboration,
        Phys. Lett. {\bf B535} (2002)430.


\bibitem{d3} C. Cawlfield, et al., CLEO Collaboration,
        Phys. Rev. {\bf D 74} (2006) 031108R.

\bibitem{d3b} M.R. Shepherd, et al., CLEO Collaboration,
        Phys. Rev. {\bf D 74} (2006) 052001.


\bibitem{d4} D. Epifanov, et al., BELLE Collaboration,
        Phys. Lett. {\bf B 654} (2007) 65.



\bibitem{d5} S. Anderson, et al., Phys. Rev. {\bf D 63}
        (2001)09001.

\bibitem{d6} B. Aubert, et al., BABAR Collaboration,
        Phys. Rev. {\bf D 76} (2007) 011102R.




\bibitem{c1} E. van Beveren, et al., Z. Phys. {\bf C 30}
        (1986)651.

\bibitem{c2} D. Aston, et al., Nucl. Phys. {\bf B 296}
        (1988)253.


\bibitem{c3} S. Ishida, et al., Prog. Theor. Phys. {\bf 98}
        (1997)621.

\bibitem{c4} D.Black, et al., Phys. Rev. {\bf D58}
        (1998)054012

\bibitem{c5} J.A. Oller, E. Oset, Phys. Rev. {\bf D 60}
        (1999)074023.

\bibitem{c6} M.J. Jamin, et al., Nucl. Phys. {\bf B 587}
        (2000)331.

\bibitem{c7} D. Bugg, Phys. Lett. {\bf B 572}
        (2003)1.

\bibitem{hanqing} H.Q.Zheng, et al., Nucl. Phys. {\bf A 733}
        (2004) 235.



\bibitem{c8} D. Lohs, Phys. Lett. {\bf B 234}
        (1990)235.

\bibitem{c9} N.A. Tornqvist, Z. Phys. {\bf C 68} (1995) 467.

\bibitem{c10} A.V. Anisovich, A.V. Sarantsev,
        Phys. Lett. {\bf B 413} (1997) 137.

\bibitem{c11} S.N. Cherry, M.R. Pennington,
        Nucl. Phys. {\bf A 688} (2001) 823.

\bibitem{pdg} Particle Data Group, Phys. Lett. {\bf B 667}
        (2008) 1.



\bibitem{besd} J.Z. Bai, et al., BES Collaboration,
        Nucl. Instrum. Methods, {\bf A} 344 (1994)319;
        J.Z. Bai, et al., BES Collaboration,
        Nucl. Instrum. Methods, {\bf A} 458 (2001)627.



\bibitem{jacob} M.~Jacob, G.C.~Wick, Ann.Phys. (NY) {\bf 7} 404 (1959).

\bibitem{chung} S.U.Chung, Phys. Rev. D57, 1998:431-442.

\bibitem{wu01}    Ning Wu and Tu-Nan Ruan, Commun. Theor. Phys.
        (Beijing, China) {\bf 35} (2001) 547.

\bibitem{wu02}   Ning Wu and Tu-Nan Ruan, Commun. Theor. Phys.
        (Beijing, China) {\bf 35} (2001) 693.

\bibitem{wu03}   Ning Wu and Tu-Nan Ruan, Commun. Theor. Phys.
        (Beijing, China) {\bf 37} (2002) 309.


\bibitem{hanqing2} H.Q. Zheng, {\it How to parameterize a
        resonance with finite width }, Talk given at
        International Symposium on Hadron Spectroscopy,
        Chiral Symmetry and Relativistic Description of
        Bound Systems, Tokyo, Japan, 24-26 Feb 2003;
        hep-ph/0304173.


\end{thebibliography}
\end{document}